\documentclass[showpacs,aps,pra,tighten,twocolumn]{revtex4-1}
\usepackage{graphicx}
\usepackage{graphics,graphicx,epsfig}

\begin{document}

\title{Trapping polarization of light in nonlinear optical fibers: An ideal Raman polarizer}

\author{Victor V. Kozlov$^{1,2,*}$, Javier Nu$\bar{\hbox{n}}$o$^3$,
Juan Diego Ania-Casta$\tilde{\hbox{n}}$\'{o}n$^3$, and Stefan Wabnitz$^{1}$}
\address{$^1$Department of Information Engineering, Universit\`{a} di
Brescia, Via Branze 38, 25123 Brescia, Italy\\
$^2$Department of Physics, St.-Petersburg State University,
Petrodvoretz, St.-Petersburg, 198504, Russia\\
$^3$Instituto de Optica, Consejo Superior de Investigaciones
Cientificas (CSIC), 28006 Madrid, Spain\\
$^*$contact information {\it victor.kozlov@email.com} }

\begin{abstract}
The main subject of this contribution is the all-optical control over the state of polarization (SOP) of light, understood as the control
over the SOP of a signal beam by the SOP of a pump beam. We will show how the possibility of such control arises naturally 
from a vectorial study of pump-probe Raman interactions in optical fibers. Most studies on the Raman effect in optical fibers 
assume a scalar model, which is only valid for high-PMD fibers (here, PMD stands for the polarization-mode dispersion). Modern 
technology enables manufacturing of low-PMD fibers, the description of which requires a full vectorial model. Within this model 
we gain full control over the SOP of the signal beam. In particular we show how the signal SOP is pulled towards and trapped 
by the pump SOP. The isotropic symmetry of the fiber is broken by the presence of the polarized pump.
This trapping effect is used in experiments for the design of new nonlinear optical devices named Raman 
polarizers. Along with the property of improved signal amplification, these devices transform an arbitrary input SOP of the signal 
beam into one and the same SOP towards the output end. This output SOP is fully controlled by the SOP of the pump beam. 
We overview the sate-of-the-art of the subject and introduce the notion of an ``ideal Raman polarizer". 
\end{abstract}

\maketitle

\section{Introduction}
Over the past few years, the possibility of utilizing the Raman effect in optical waveguides as the basis for the development of non-linear polarizers
has opened the way to an interesting range of potential applications, such as multi-channel repolarization in optical fibers, enhanced amplification 
and even the possibility of developing silicon-based Raman polarizers \cite{ourFutureOL,our1,our2,Sergeyev,silicon}.

Raman-based polarization attraction falls into a broad class of potentially game-changing effects related to light-by-light control in optical waveguides. 
Models for such control are essentially nonlinear and usually imply the use of a high-intensity beam to modify the properties of the medium (for 
instance its refractive index or absorption coefficient) such that propagation of a weaker probe beam through the nonlinearly modified medium is 
affected in a substantial and controllable way. The possibility of achieving nonlinear polarization control is rooted in soliton theory, namely in 
conservative structures such as the polarization domain wall solitons \cite{Zakharov87,Daino,malomed,Pitois0,Pitois1,Stefan1}. However conclusions extracted from soliton theories  
involving a medium of infinite extension can be misleading for counterpropagating waves in a medium of finite length. In this case the presence of boundary conditions may lead to solitons with a finite lifetime \cite{LMP}.
In such situation, other so-called polarization attractors representing the unique distribution of SOPs of the two beams inside the medium play a key role 
in the process of trapping polarization of light \cite{LMP,DominiqueJOSAB2012}. 

Different mechanisms such as photorefractive 
two-beam coupling \cite{Heebner} or Kerr nonlinearity \cite{Pitois1,Pitois2} have, over the years, proven to be capable of producing nonlinear 
polarization attraction. In their initial demonstrations, all of these methods were subject to limitations 
in their application in telecommunication links: their response time, in the case of photorefractive materials, or the requirement of extremely high beam 
powers. Only recently results of practical relevance have emerged, with non-conservative schemes based on stimulated 
Raman \cite{martinelli} or Brillouin scattering \cite{Zadok}, as well as the first low-power lossless polarizer, consisting of a 20 km randomly weakly
birefringent fiber pumped by an incoherent counter-propagating beam \cite{Fatome}.
 
As mentioned above, here we will focus on the particular and very promising case of Raman polarizers, in which the pump and signal beam propagate
through a Raman-active medium. By way of interacting with this medium, the pump beam induces a phonon-mediated gain for a frequency down-shifted
(Stokes) signal beam. The signal beam, co- or counter-propagating with the pump beam, is then gradually amplified. This amplification mechanism lies
at the heart of Raman amplifiers. One degree of control exerted by the pump beam over the signal beam is the total gain experienced by the signal from
input to the output. This degree of control is well studied in literature and widely used in practice. Much less known is another degree of control -- over
the state of polarization (SOP) of the signal beam. The main subject of this study are {\it polarization-sensitive} Raman amplifiers, in which 
polarization-dependent gain (PDG), an intrinsic characteristic of the Raman effect which is usually considered an undesirable feature in amplification
applications, can be turned into an advantage by selectively amplifying only one polarization mode of the input beam.

Signal and pump fields considered in this study are continuous waves (CW) or relatively long pulses, such that the response of 
the Raman-active medium is virtually instantaneous, and as such it is described by the instantaneous dissipative cubic nonlinearity. 
Mostly, our theoretical study is developed for silica single-mode fibers, though extensions to other Raman-active media, such as 
silicon are also possible \cite{silicon}. We shall demonstrate how polarization-sensitive Raman amplifiers operate in the regime
of Raman polarizers. These Raman polarizers are devices that along with the function of amplification of light, also re-polarize the
beam: the SOP of the outcoming signal beam is defined by the SOP of the pump beam, independently of what SOP the signal beam
had at the input. In other words, the signal SOP is attracted (trapped) by the pump SOP. By changing the polarization of the pump 
we thereby change the signal SOP.  In this way we exercise an all-optical control over the signal SOP. 

In this chapter we will present the theory of Raman polarizers with an emphasis on randomly birefringent fibers, such as the ones used in the telecom 
industry. We shall identify the conditions that are necessary for a traditional Raman amplifier to function as Raman polarizer, and characterize its 
performance. 

\section{Model}
In short, we shall consider the simultaneous propagation of two beams in a Raman-active medium. In our case the Raman active 
medium is a few kilometers long span of a telecom fiber. The fiber is linearly birefringent, and also characterized by both conservative and dissipative
cubic nonlinearities. The main feature that makes our theory different from most previous studies on fiber-optic Raman amplifiers is its vectorial nature.
Thus, we carefully consider the propagation dynamics of two polarization components of each of the two beams. In total, the number of field
components is four, and they all interact with each other via cubic nonlinearity. The first vectorial theory of Raman effect in randomly birefringent optical
fibers was developed by Lin and Agrawal in Ref.~\cite{agrawal} and applied to the regime of interaction characteristic to what we call here ``standard
Raman amplifiers". Here we are interested in a totally different regime, namely the regime of Raman polarizer. The difference between the two regimes
is explained below, in the beginning of section II.

We start from the equation of motion for the signal field, written  for the two-component field vector $U_s=(u_{sx},\, u_{sy})^T$, 
where $u_{sx}$ and $u_{sy}$ are the amplitudes of the normal polarization modes ${\textbf e}_x$ and ${\textbf e}_y$ of the fiber: 
${\textbf U}_s=u_{sx}{\textbf e}_x+u_{sy}{\textbf e}_y$. This equation is derived under the (as usual for nonlinear optics) 
unidirectional and slowly varying approximations, see for instance \cite{agrawal,Menyuk}, and reads
\begin{eqnarray}
&& i\partial_z U_s
+i\beta^\prime (\omega_p)\partial_t U_s
+\Delta B(\omega_s)U_s
\nonumber\\
&& +\gamma_{ss}\left[
\frac{2}{3}(U_s^*\cdot U_s)U_s+\frac{1}{3}(U_s\cdot U_s)U_s^*
\right]
\nonumber\\
&& + \frac{2}{3}\gamma_{sp}\left[
(U_p^*\cdot U_p)U_s+(U_p\cdot U_s)U_p^*+(U_s\cdot U_p^*)U_p
\right]
\nonumber\\
&& +i\epsilon_s g (U_p^*\cdot U_s)U_p=0\, .
\label{1}
\end{eqnarray}
A similar equation (with indices $p$ and $s$ interchanged) arises for the pump beam, which is characterized by the field 
vector $U_p$. Here $\gamma_{ss}$ and $\gamma_{sp}$ are self- and cross-modulation coefficients, whose values 
depend on frequency, and therefore in principle are different for the signal and pump beams. They are equal to the 
frequency-dependent Kerr coefficient of the fiber. For simplicity we assume $\gamma_{ss}=\gamma_{pp}=\gamma_{ps}\equiv \gamma$.
$\beta^\prime (\omega_{p,s})$ is the inverse group velocity of the pump/signal beam. 
$\epsilon_s=1$, $\epsilon_p=-\omega_{s}/\omega_{p}$. $\Delta B(\omega_{p,s})$ is the birefringence tensor.  For a linearly 
birefringent fiber it takes the form $\Delta B (\omega_{p,s})=\Delta\beta (\omega_{p,s})\left(\cos\theta \sigma_3+\sin\theta\sigma_1\right)$, 
where $\Delta\beta (\omega_{p,s})$ is the value of birefringence at frequency $\omega_{p,s}$, and $\theta$ the angle of orientation of 
the axis of the birefringence with respect to the reference frame defined by polarization modes ${\textbf e}_x$ and ${\textbf e}_y$. 
$\sigma_3$ and $\sigma_1$ are the usual Pauli matrices.

The orientation angle $\theta$ is randomly varying in fibers. In principle, the magnitude of the birefringence $\Delta\beta$ also varies 
stochastically. However, as noticed in Ref.~\cite{wai_menyuk}, the two approaches, one in which $\theta$ is the only stochastic 
variable, and the second, where both $\theta$ and $\Delta\beta$ are stochastic variables, produce nearly identical results. Thus, here we shall develop 
our theory by assuming the single stochastic variable $\theta$. Our theory can be seen as a generalization of the one beam linearly birefringent theory of 
Wai and Menyuk from Ref.~\cite{wai_menyuk} to the case of two beams interacting via the Kerr and Raman nonlinearity in a fiber. The angle $\theta$ is 
driven by a white noise process $\partial_z\theta =g_\theta (z)$,  where $\langle g_\theta (z)\rangle =0$ and 
$\langle g_\theta (z) g_\theta (z^\prime )\rangle = 2L_c^{-1}\delta (z-z^\prime )$. Here $L_c$ is the correlation length, that characterizes the typical 
distance at which $\theta$ changes randomly. 

Details of the theory are presented in Refs.~\cite{our1,our2}. Here we quickly drive through the major steps of this theory and show how to obtain the final
result -- a set of four coupled first-order ordinary differential equations, one equation for each polarization component of two beams. It is instructive to
present these four equations as two vectorial equations for the Stokes vectors of the pump and signal beams.  Each Stokes vector has three
components. Namely, the pump Stokes vector ${\mathbf S}^{(p)}=(S_1^{(p)},\, S_2^{(p)},\, S_3^{(p)})$ has components 
$S_1^{(p)} =\Psi_{p1}^*\Psi_{p2} +\Psi_{p1} \Psi_{p2}^*$, $S_2^{(p)} = i\left(\Psi_{p1}^*\Psi_{p2} -\Psi_{p1} \Psi_{p2}^*\right)$, 
$S_3^{(p)} =\vert\Psi_{p1}\vert^2-\vert\Psi_{p2}\vert^2$, and power $S_0^{(p)}=\vert {\mathbf S}^{(p)}\vert$. Similar expressions define the 
signal Stokes vector ${\mathbf S}^{(s)}$. Here the field vector $\Psi$ is related to the original Jones field vector $U$ by the relation 
$\Psi_{p,s}=T_{p,s}U_{p,s}$, where the $2\times 2$ matrices $T_{p,s}$ with elements
\begin{eqnarray}
&& T_p(z)=\left(
\begin{array}{cc}
a_1 & a_2\\
-a_2^* & a_1^*
\end{array}\right)\, ,
\label{5JLT}\\
&& T_s(z)=\left(
\begin{array}{cc}
b_1 & b_2\\
-b_2^* & b_1^*
\end{array}\right)\, .
\label{6JLT}
\end{eqnarray}
obey the stochastic differential equations
\begin{eqnarray}
\pm i\partial_z T_p+\Delta B(\omega_p) T_p=0\, ,
\label{7JLT}\\
i\partial_z T_s+\Delta B(\omega_s) T_s=0\, ,
\label{7_1JLT}
\end{eqnarray}
where plus (minus) sign stands for the co- (counter-) propagating regime of propagation of the two beams, and
\begin{eqnarray}
&& \Delta B(\omega_p)=
\left(
\begin{array}{cc}
\Delta\beta (\omega_p) & \mp\frac{i}{2}\theta_z \\
\pm\frac{i}{2}\theta_z & -\Delta\beta (\omega_p)
\end{array}\right)\, ,
\label{4JLT}\\
&& \Delta B(\omega_s)=
\left(
\begin{array}{cc}
\Delta\beta (\omega_s) & -\frac{i}{2}\theta_z \\
\frac{i}{2}\theta_z & -\Delta\beta (\omega_s)
\end{array}\right)\, .
\label{4_1JLT}
\end{eqnarray}
Here, $\theta_z$ is the derivative of $\theta$ with respect to $z$. It is different from zero owing to the random changes of orientation of the
birefringence axes. Now the polarization components of each beam appear to be defined with respect to the local birefringence axes, while these 
axes rotate stochastically along the fiber length driven by the noise source $g_\theta (z)$.  

These transformations eliminate the birefringence
terms from the equations of motion of $\Psi_p$ and $\Psi_s$
and bring about a vast number of cubic terms composed
of different combinations of $\Psi_{p1}$, $\Psi_{p2}$,
$\Psi_{s1}$, $\Psi_{s2}$ and their complex conjugates.
Factors in front of these terms are products of two
coefficients of the form $u_mu_n$, or $u_m^*u_n$,
or $u_m^*u_n^*$, where $m,n=1,\,\dots ,\, 14$.
Products with $m=n$ we shall call self-products,
while with $m\ne n$ cross-products. Here,
$u_1=\vert a_1\vert^2-\vert a_2\vert^2$,
$u_2=-(a_1a_2+a_1^*a_2^*)$,
$u_3=i(a_1a_2-a_1^*a_2^*)$,
$u_4=2a_1a_2^*$, $u_5=a_1^2-{a_2^*}^2$,
$u_6=-i(a_1^2+{a_2^*}^2)$,
$u_7=a_1^*b_1-a_2b_2^*$,
$u_8=-(b_1a_2+b_2^*a_1^*)$,
$u_9=i(b_1a_2-a_1^*b_2^*)$,
$u_{10}=-i(a_1^*b_1+a_2b_2^*)$,
$u_{11}=a_1b_2^*+b_1a_2^*$,
$u_{12}=a_1b_1-a_2^*b_2^*$,
$u_{13}=-i(a_1b_1+a_2^*b_2^*)$,
$u_{14}=i(a_1b_2^*-a_2^*b_1)$.

In the thus obtained equations of motion for
$\Psi_p$ and $\Psi_s$ we perform the ensemble
average (over different realizations of the random process
which describes linear birefringence).
Thus, we write $\langle u_m u_n\rangle$
instead of $u_mu_n$. This change holds true
only in the limit when the stochastic variations are
faster than the nonlinear beam evolution. This is exactly
the place in the derivation where our single
approximation comes into play. At this point we
also need to apply the ergodic theorem
\begin{equation}
\langle f\rangle =
\lim_{z\to\infty}\frac{1}{z}\int_0^zdz^\prime\, f(z^\prime )\, .
\label{8JLT}
\end{equation}
Our goal is to calculate ensemble averages of all
necessary self- and cross-products: in this way we may
complete the derivation of the differential equations for
$\Psi_p$ and $\Psi_s$.

The equations of motion for $u_n$ with $n=1,\,\dots ,\, 14$
can be easily formulated basing ourselves on equations (\ref{7JLT})
and (\ref{7_1JLT}). As these equations are linear, in order to find an
ensemble average of any function of these coefficients (in
our case pair products) we need to construct a
generator. We refer to the Appendix in
Ref.~\cite{wai_menyuk} for details of this procedure,
and only give here the final result. With this generator
we are able to formulate the equations of motion for the
ensemble averages of the products of the coefficients.
Thus the solutions to the equations of motion
\begin{eqnarray}
&& \partial_z G_1=
-2L_c^{-1}(G_1-G_2)\, ,
\label{9JLT}\\
&& \partial_z G_2=
2L_c^{-1}(G_1-G_2)\mp 4\Delta\beta (\omega_p)G_4 \, ,
\label{10JLT}\\
&& \partial_z G_3=
\pm 4\Delta\beta (\omega_p)G_4 \, ,
\label{11JLT}\\
&& \partial_z G_4=
-L_c^{-1}G_4\pm 2\Delta\beta (\omega_p)(G_2-G_3)
\label{12JLT}
\end{eqnarray}
yield the result for the self-products
$\{ \langle u_1^2\rangle ,\, \langle  u_2^2\rangle ,\, \langle u_3^2\rangle\}$,

\noindent
$\{ \langle \hbox{Re}^2(u_4)\rangle ,\, \langle\hbox{Re}^2(u_5)\rangle ,\,
\langle \hbox{Re}^2(u_6)\rangle\}$,

\noindent
and $\{ \langle \hbox{Im}^2(u_4)\rangle ,\,
\langle\hbox{Im}^2(u_5)\rangle ,\, \langle \hbox{Im}^2(u_6)\rangle\}$, if we
associate them with $\{ G_1,\, G_2,\, G_3\}$ with initial conditions
given as $(1,\, 0,\, 0)$, $(0,\, 1,\, 0)$, and $(0,\, 0,\, 1)$, respectively.

The remaining self-products
$\{ \langle \hbox{Re}^2(u_7)\rangle ,\, \langle\hbox{Re}^2(u_8)\rangle ,\,
\langle \hbox{Re}^2(u_9),\, \langle\hbox{Re}^2(u_{10})\rangle\}$,

\noindent
$\{ \langle \hbox{Im}^2(u_7)\rangle ,\, \langle\hbox{Im}^2(u_8)\rangle ,\,
\langle \hbox{Im}^2(u_9),\, \langle\hbox{Im}^2(u_{10})\rangle\}$,

\noindent
$\{ \langle \hbox{Re}^2(u_{11})\rangle ,\, \langle\hbox{Re}^2(u_{12})\rangle ,\,
\langle \hbox{Re}^2(u_{13}),\, \langle\hbox{Re}^2(u_{14})\rangle\}$,

\noindent
and $\{ \langle \hbox{Im}^2(u_{11})\rangle ,\, \langle\hbox{Im}^2(u_{12})\rangle ,\,
\langle \hbox{Im}^2(u_{13}),\, \langle\hbox{Im}^2(u_{14})\rangle\}$,
can be found from the equations
\begin{eqnarray}
&& \partial_z G_1=-2L_c^{-1}(G_1-G_2)
+2\Delta_\pm G_5\, ,
\label{13JLT}\\
&& \partial_z G_2=2L_c^{-1}(G_1-G_2)
-2\Delta_\pm G_6\, ,
\label{14JLT}\\
&& \partial_z G_3=
2\Delta_\pm G_6\, ,
\label{15JLT}\\
&& \partial_z G_4=
-2\Delta_\pm G_5\, ,
\label{16JLT}\\
&& \partial_z G_5=
\Delta_\pm (G_4-G_1)
-L_c^{-1}G_5\, ,
\label{17JLT}\\
&& \partial_z G_6=
\Delta_\pm (G_2-G_3)
-L_c^{-1}G_6\, ,
\label{18JLT}
\end{eqnarray}
when we associate them with
$\{ G_1,\, G_2,\, G_3,\, G_4\}$, with initial conditions
as $(1,\, 0,\, 0,\, 0)$, $(0,\, 0,\, 0,\, 1)$, $(0,\, 1,\, 0,\, 0)$,
and $(0,\, 0,\, 1,\, 0)$, respectively.  Here
$\Delta_\pm\equiv\left[\pm\Delta\beta (\omega_p)-\Delta\beta (\omega_s)\right]$.

In order to find the cross-products we constructed appropriate
generators and found that all the cross-products that are of
interest to us turn out to be equal to zero. Similarly, terms of the form
$\hbox{Re}(u_n)\hbox{Im}(u_n)$ also vanish. Thus, many
SPM, XPM, and Raman terms in the final equations of motion
disappear. The thus found equations of motion for the fields
are conveniently formulated in Stokes space. They read as
\begin{eqnarray}
&& \left(\pm\partial_z
+\beta^\prime (\omega_p)\partial_t\right)
{\mathbf S}^{(p)} =
\nonumber\\
&& \gamma 
\left({\mathbf S}^{(p)}\times J_s^{(p)}(z) {\mathbf S}^{(p)}
+{\mathbf S}^{(p)}\times J_x(z) {\mathbf S}^{(s)}\right)
\nonumber\\
&& +\epsilon_p(g/2)\left(S_{0}^{(s)}J_{R0}{\mathbf S}^{(p)}
+S_0^{(p)} J_R(z){\mathbf S}^{(s)}
\right)\, ,
\label{19JLT}\\
&& \left(\partial_z
+\beta^\prime (\omega_s)\partial_t\right)
{\mathbf S}^{(s)}=
\nonumber\\
&& \gamma 
\left({\mathbf S}^{(s)}\times J_s^{(s)}(z) {\mathbf S}^{(s)}
+{\mathbf S}^{(s)}\times J_x(z) {\mathbf S}^{(p)}\right)
\nonumber\\
&& +(g/2)\left(S_{0}^{(p)}J_{R0}{\mathbf S}^{(s)}
+S_0^{(s)}J_R(z) {\mathbf S}^{(p)}\right)\, .
\label{20JLT}
\end{eqnarray}

Matrices in equations (\ref{19JLT}) and (\ref{20JLT})
are all diagonal with elements

\noindent
$J_R=\hbox{diag}(J_{R1},\, J_{R2},\, J_{R3})$,
$J_x=\hbox{diag}(J_{x1},\, J_{x2},\, J_{x3})$,
$J_s=\hbox{diag}(J_{s1},\, J_{s2},\, J_{s3})$.
These elements are different for the counter-propagating
and the co-propagating interaction geometries.

In order to complete our theory, we need to express
all elements in these matrices in terms of ensemble
averages of self-products:
\begin{eqnarray}
&& J_{R1}=\langle \hbox{Re}(u_{14}^2-u_{10}^2)\rangle\, ,
\label{26JLT}\\
&& J_{R2}=-\langle \hbox{Re}(u_{14}^2+u_{10}^2)\rangle\, ,
\label{27JLT}\\
&& J_{R3}=-\langle \vert u_{14}\vert^2-\vert u_{10}\vert^2\rangle\, ,
\label{28JLT}\\
&& J_{x1}=\frac{2}{3}\langle \hbox{Re}(u_{10}^2
+u_{13}^2-u_9^2-u_{14}^2)\rangle\, ,
\label{29JLT}\\
&& J_{x2}=\frac{2}{3}\langle \hbox{Re}(u_{10}^2
+u_{14}^2-u_9^2-u_{13}^2)\rangle\, ,
\label{30JLT}\\
&& J_{x3}=\frac{2}{3}\langle \vert u_{9}\vert^2+\vert u_{14}\vert^2
-\vert u_{13}\vert^2-\vert u_{10}\vert^2\rangle\, ,
\label{31JLT}\\
&& J_{s1}=\frac{1}{3}\langle \hbox{Re}(u_{6}^2)\rangle\, ,
\label{32JLT}\\
&& J_{s2}=-\frac{1}{3}\langle \hbox{Re}(u_{6}^2)\rangle\, ,
\label{33JLT}\\
&& J_{s3}=\frac{1}{3}\left[3\langle u_3^2\rangle -1\right]\, ,
\label{34JLT}
\end{eqnarray}
and also $J_{R0}=\langle \vert u_{10}\vert^2+\vert u_{14}\vert^2\rangle$.
Note that our model reduces to the one-beam theory of Wai and
Menyuk when the coefficients $u_7$ through $u_{14}$ are set to zero.

The Stokes representation is particularly appealing in the context of the problem that we are considering. As we are interested in the 
polarization properties of the outcoming signal beam, the Stokes vector quite clearly presents the polarization vector on the Poincar\'{e} sphere. The 
evolution of the Stokes vector draws a trajectory of its tip on the sphere. Another quantity of interest is the degree of polarization
(DOP). In those cases where we are dealing with an ensemble of beams, the DOP characterizes the length of the average Stokes vector. Here again 
the Stokes representation appears to be rather useful. 

Thus, the equation of motion for the Stokes vector of a CW signal beam is
\begin{eqnarray}
\partial_z {\mathbf S}^{(s)} & = & \gamma {\mathbf S}^{(s)}\times J_s(z) {\mathbf S}^{(s)}+
                                                      \gamma {\mathbf S}^{(s)}\times J_x(z) {\mathbf S}^{(p)}
\nonumber\\
                                           & + & \epsilon_{p}(g/2)\left[ S_0^{(p)} {\mathbf S}^{(s)} +S_0^{(s)} J_R(z) {\mathbf S}^{(p)}\right]\, .
\label{3}
\end{eqnarray}
(With $J_{R0}=1$, which is the case for all situations considered below.)
Here $J_s$ is the self-polarization modulation (SPolM) tensor, $J_x$ -- cross-polarization (XPolM) modulation tensor, $J_R$ -- Raman 
tensor. All they  are diagonal. Elements of these tensors are dependent on the magnitude of the birefringence both at signal and pump 
carrier frequencies, that is on the beat lengths $L_B(\omega_{s})$ and $L_B(\omega_{p})$, and also on the correlation length $L_c$. 
All these three lengths do not exceed $100$~m in conventional telecom fibers. The physical meaning of each tensor follows from its 
definition. Thus, the SPolM tensor defines how two polarization components belonging to the same beam interact in the Kerr medium, and thereby 
rotate the Stokes vector. The XPolM tensor has similar meaning, but now the rotation is due to the interaction of polarization components belonging to 
different beams. Finally, the Raman tensor defines polarization-sensitive amplification of 
amplifier. This tensor is of particular importance to us. For instance, when all elements of this tensor vanish, the Raman 
amplifier becomes insensitive to the SOP of the pump beam, so that we are dealing essentially with a {\it scalar} model. Conversely, when 
the diagonal elements of the Raman tensor have appreciable values, then the theory must be necessary {\it vectorial}. 

\begin{figure}
\begin{center}
\includegraphics[scale=0.5]{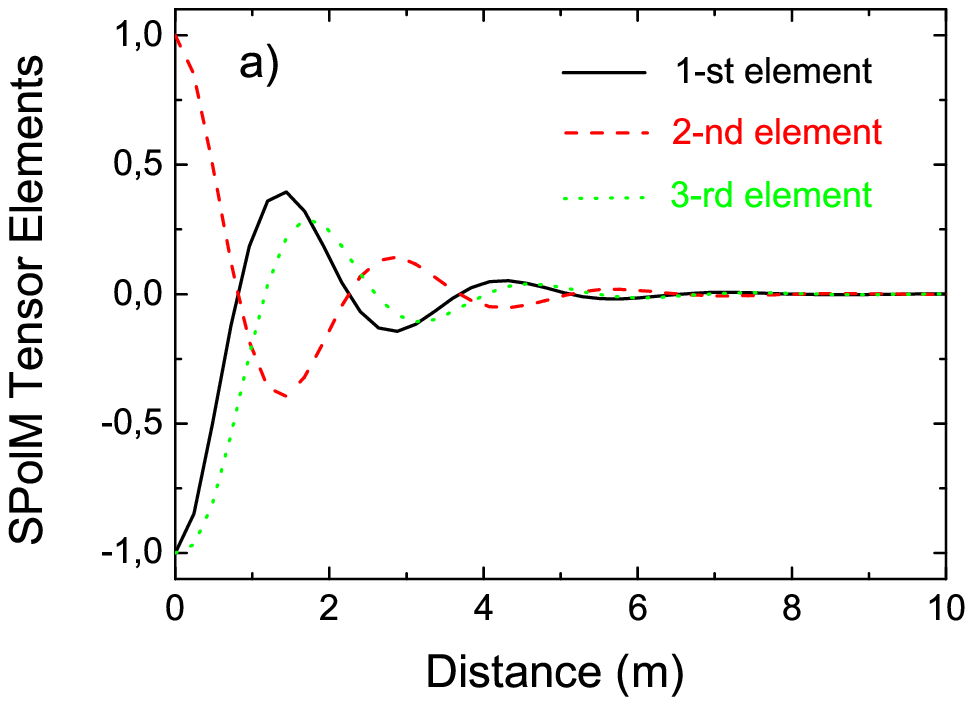}
\includegraphics[scale=0.5]{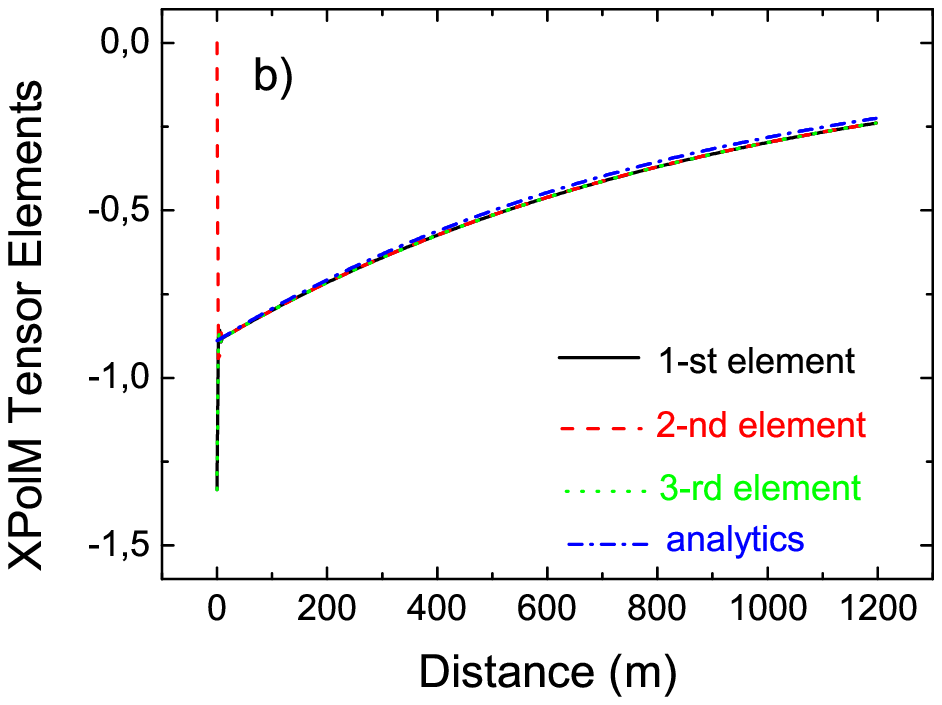}
\includegraphics[scale=0.5]{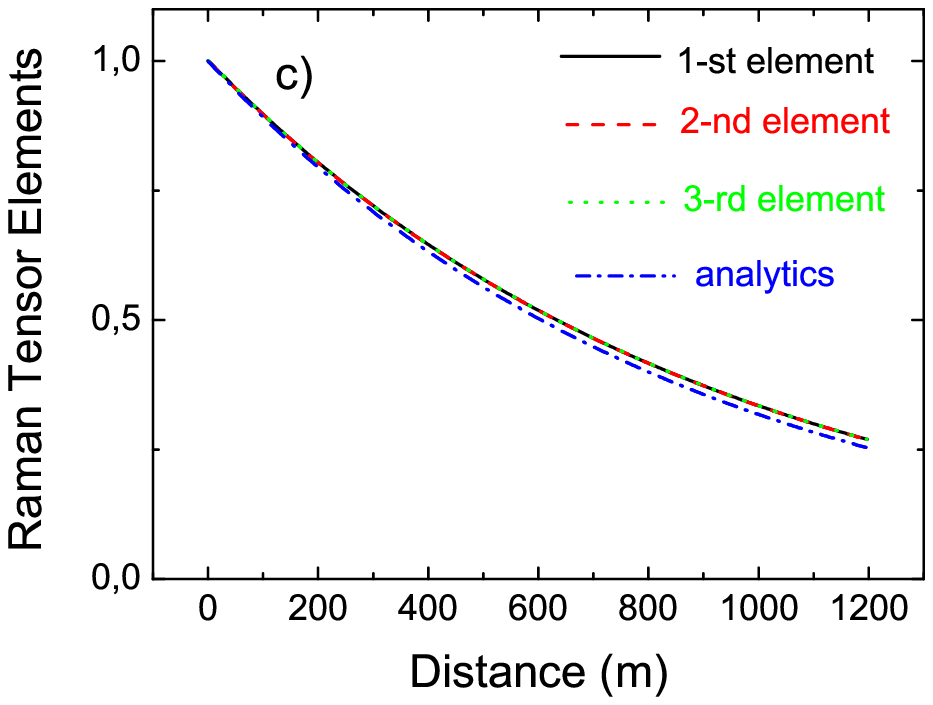}
\end{center}
\caption{Three diagonal elements of the (a) SPolM, (b) XPolM, and (c) Raman tensors. In figures (b,c) all three curves visually coincide;
the blue curve is the analytical result  showing the exponential decay: $\propto \exp (-z/L_d)$. Parameters are: $L_c=1$~m, 
$L_B(\omega_s)=10$~m, $\omega_p-\omega_s=13.2$~THz, $\lambda_s=1.55$~$\mu$m, and $\lambda_p=1.45$~$\mu$m. The PMD
diffusion length is $L_d=870$~m. Note that a brief transient in Fig.~\ref{ris1}(b) is not resolved on the chosen scale.
}
\label{ris1}
\end{figure}

Certainly, the evolution of the Stokes vector sensitively depends on how the elements of these tensors evolve with distance. In order to find their
dynamics it is necessary to solve the set of linear ordinary differential equations which is given above, see also Refs.~\cite{our1,our2}. Instead of writing
them down here, we present their approximate analytical solutions. Fig.~\ref{ris1}(a,b,c) shows how well these analytical  solutions reproduce the 
exact situation. Fig.~\ref{ris1}(a) shows that the elements of the SPolM tensor drop very fast and already vanish within the first $10$~m of the 
fiber. Given, that the length of the Raman amplifier exceeds $1\div 2$~km, we can safely set
\begin{equation}
J_s=\hbox{diag}(0,\, 0,\, 0)\, .
\label{4}
\end{equation}
The elements of the other two tensors also deceases with distance, however much slower, namely as
\begin{eqnarray}
&&  J_x=-\frac{8}{9}\hbox{diag}(1,\, 1,\, 1)\exp (-z/L_d)\, ,
\label{5}\\
&&  J_R=\hbox{diag}(1,\, 1,\, 1)\exp (-z/L_d)\, .
\label{6}
\end{eqnarray}
As demonstrated in Fig.~\ref{ris1}(b,c) the decay distance is indeed determined by the characteristic length $L_d$, which is called the PMD diffusion 
length: $L_d^{-1}=\frac{1}{3}(D_p\Delta\omega )^2$, where $D_p=2\sqrt{2}\pi \sqrt{L_c}/(L_B\omega_p)$ is the PMD coefficient, \cite{wai_menyuk}, and 
throughout the paper $\Delta\omega =\omega_p-\omega_s$ is taken 
to be equal to the Raman shift $\Delta\omega_R$ in the germanium-doped silica fibers, that is $13.2$~THz. The theory that we are 
developing here is strictly valid only in two limits -- the limit which we call here Manakov limit ($L_{NL},\, L_R\ll L_d$) and diffusion limit  
($L_{NL},\, L_R\gg L_d$), where $L_{NL}$ is nonlinear length, and $L_R$ characteristic amplification length.

\section{Raman amplifiers versus Raman polarizers}
Raman amplifiers, which we call here standard Raman amplifiers, operate in the diffusion limit, as they are based on fibers with large PMD coefficients.
Thus, for $D_p=0.2$~ps$/\sqrt{\hbox{km}}$ and $\Delta\omega =\Delta\omega_R=13.2$~THz, the PMD diffusion length $L_d$ is less than 
$10$~m. Taking into account that standard Raman amplifiers are $10$ or more kilometers long, the contribution of the
polarization-dependent gain (second term in brackets in Eq.~(\ref{3})) to the total gain (both terms in brackets in Eq.~(\ref{3}) taken
together) is totally negligible. The model equation for the signal beam is then
\begin{eqnarray}
\partial_z {\mathbf S}^{(s)}  = (g/2) S_0^{(p)} {\mathbf S}^{(s)}\, .
\label{7}
\end{eqnarray}
Thus, each component of the Stokes vector is amplified independently and equally with the other components. For such Raman amplifier
there is no preferentially amplified polarization mode. The model is essentially a {\it scalar} one.

A different situation arises in the Manakov limit. For PMD coefficients less than $0.02$~ps$/\sqrt{\hbox{km}}$, the PMD diffusion length becomes
greater than $1$~km. In this case we can write the model equation for the signal Stokes vector in the form
\begin{eqnarray}
\partial_z {\mathbf S}^{(s)} & = & -\bar\gamma {\mathbf S}^{(s)}\times {\mathbf S}^{(p)}
\nonumber\\
                                           & + & (g/2)\left[ S_0^{(p)} {\mathbf S}^{(s)} +S_0^{(s)} {\mathbf S}^{(p)}\right]\, ,
\label{8}
\end{eqnarray}
with $\bar\gamma =\frac{8}{9}\gamma$. In this limit ($L_d\to\infty$) we deal with an ideal Raman polarizer. The equation above includes two
contributions. The XPolM contribution is a cross-phase modulation (XPM) part of the Manakov equation, in which the factor of $\frac{8}{9}$ appears as
the result of averaging of fast stochastic polarization dynamics of each Stokes vector. Quite to the contrary, the Raman contribution appears exactly 
as in the case of isotropic fibers (i.e. in absence of the birefringence, and its stochasticity), because the mutual polarization scrambling of the relative
orientations of the pump and Stokes vectors is very inefficient when the PMD diffusion length $L_d$ is long. In other words, Raman amplification 
is insensitive to the absolute orientation of the individual SOPs of the signal and pump beams in the laboratory frame. It is however sensitive to their {\it
mutual} orientation. In the case of standard Raman amplifiers, the signal Stokes vector rotates rapidly around the pump Stokes vector, and therefore on
average ``feels" no polarization dependence. In the case of Raman polarizers, still the two vectors stochastically rotate in the laboratory frame, but they
do it now in unison, so that their mutual orientation is almost ``frozen". 

\section{An ideal Raman polarizer}
As characteristic to isotropic fibers, the signal experiences maximal gain when its Stokes vector is aligned along the pump Stokes vector. To show
this we can choose (without loss of generality) the pump Stokes vector be aligned along its first component: 
${\mathbf S}^{(p)}=S_0^{(p)}(1,\, 0,\, 0)$. Then, we may write for the signal first Stokes component:
\begin{equation}
\partial_z S_1^{(s)} = (g/2)S_0^{(p)} \left[S_0^{(s)}+ S_1^{(s)}\right]\, .
\label{9}
\end{equation}
If initially the signal Stokes vector is also aligned with its first component, then the signal amplification coefficient is $g$. This value should be contrasted to
the value of $g/2$, which is characteristic to standard Raman amplifiers, see Eq.~(\ref{7}). 

The other two components of the signal Stokes vector are amplified less efficiently than the first component. Indeed, their equations of motion are:
\begin{eqnarray}
\partial_z S_2^{(s)} = -\bar\gamma S_0^{(p)}S_3^{(s)}+(g/2)S_0^{(p)} S_2^{(s)}\, ,
\label{10}\\
\partial_z S_3^{(s)} = \bar\gamma S_0^{(p)}S_2^{(s)}+(g/2)S_0^{(p)} S_3^{(s)}\, .
\label{11}
\end{eqnarray}
Here, the gain is only $g/2$.

The observations derived from Eqs.~(\ref{9})-(\ref{11}) explain the ability of a Raman polarizer to re-polarize light. They demonstrate that only
the Stokes component of the signal aligned with the pump Stokes vector is dominantly amplified. In a high-gain Raman amplifier, the difference
in gain for polarization components may become so large that the polarization of the outcoming beam is almost perfectly aligned with the pump SOP. 
This effect of alignment is called polarization attraction, or polarization trapping. Shortly, we shall quantify effect of the polarization attraction
in terms of the DOP, the so-called alignment parameter, and some other parameters, while now we comment on the output SOP of the outcoming signal 
beam measured with  respect of the laboratory frame. 

As we have seen, the Raman tensor decays as the distance grows larger, see Eq.~(\ref{6}). Therefore, it is preferable to decrease the total fiber
length at the price of increasing the pump power. Indeed, the first proof-of-principle experiment reported in Ref.~\cite{martinelli} by Martinelli {\it et. al.}, 
was carried out with a dispersion-shifted fiber of only $2,1$~km, and an average pump power as high as $2.2$~W.

Most theories of Raman polarizers reported so far, see Refs.~\cite{martinelli,our1,our2,padova1,padova2}, are based on numerical simulations of
the stochastic equations which properly take into account the randomness of the fiber birefringence (the total fiber span is divided into short segments,
with each segment extended over one correlation length; the orientation of the birefringence axes is fixed within each segment, while it varies randomly
when going from one segment to the next one). Such an approach is indeed necessary when
the PMD diffusion length is comparable with the amplification length, a case which is in the middle between the Manakov limit 
and the diffusion limit. In this case, the Raman polarizer has non-optimal performances, yielding a DOP which is significantly below unity. So, this case 
is not advantageous in practice. In order to realize a ``good" Raman polarizer, one should choose to work in the Manakov limit. As we have indicated 
above, working in this limit allows us to get analytical and physically transparent results. In the next section we shall continue to work with ideal 
Raman polarizers and provide an even deeper analytical insight. 

\section{Evaluation of the performance of a Raman polarizer}
Equations (\ref{9})-(\ref{11}) can be solved analytically. We shall limit ourselves to the undepleted pump approximation, so that the pump power
$P\equiv S_0^{(p)}(z)=$const. Our model does not include linear losses in the fiber, because we have chosen to work with relatively short fiber
spans, for which losses are relatively small. If necessary, the losses can be included, though analytics will become less transparent.
Solutions to Eqs.~(\ref{9})-(\ref{11}) are:
\begin{eqnarray}
&& S_0^{(s)}(z)=\frac{1}{2}\left[ S_{0}^{(s)}(0)-S_{1}^{(s)}(0)\right]
\nonumber\\
&& +\frac{1}{2}\left[ S_{0}^{(s)}(0)+S_{1}^{(s)}(0)\right]\hbox{e}^{gPz}\, ,
\label{12}\\
&& S_1^{(s)}(z)=-\frac{1}{2}\left[ S_{0}^{(s)}(0)-S_{1}^{(s)}(0)\right]
\nonumber\\
&& +\frac{1}{2}\left[ S_{0}^{(s)}(0)+S_{1}^{(s)}(0)\right]\hbox{e}^{gPz}\, ,
\label{13}\\
&& S_2^{(s)}(z)=\Big[ S_{2}^{(s)}(0)\cos (\bar\gamma Pz)
\nonumber\\
&& -S_3^{(s)}(0)\sin (\bar\gamma Pz)\Big]\hbox{e}^{\frac{1}{2}gPz}\, ,
\label{14}\\
&& S_3^{(s)}(z)=\Big[ S_{2}^{(s)}(0)\sin (\bar\gamma Pz)
\nonumber\\
&& +S_3^{(s)}(0)\cos (\bar\gamma Pz)\Big]\hbox{e}^{\frac{1}{2}gPz}\, .
\label{15}
\end{eqnarray}

We are interested in the statistical properties of a Raman polarizer. All quantities of interest can be derived from the above-written solutions. First of all,
we shall calculate the mean quantities. The immediate questions are -- what is the SOP of the outcoming signal beam and how well the beam is
polarized? In order to find an answer to the first question we need to simply get an average of Eqs.~(\ref{12})-(\ref{15}) given the statistics of the incoming
light. We assume that the signal is initially unpolarized, so that $\langle S_1^{(s)}(0)\rangle =   \langle S_2^{(s)}(0)\rangle = \langle S_3^{(s)}(0)\rangle = 0$.
Then, at $z=L$, where $L$ is the total length of the fiber, we get
\begin{eqnarray}
\langle S_0^{(s)}(L)\rangle = \frac{1}{2} S_{0}^{(s)}(0) \left[ 1+\exp (gPL)\right]\, ,
\label{16}\\
\langle S_1^{(s)}(L)\rangle = \frac{1}{2} S_{0}^{(s)}(0) \left[ -1+\exp (gPL)\right]\, ,
\label{17}\\
\langle S_2^{(s)}(L)\rangle = 0\, ,
\label{18}\\
\langle S_3^{(s)}(L)\rangle = 0\, .
\label{19}
\end{eqnarray}
So, the signal SOP at the output is aligned with the pump SOP.  The degree of alignment  is characterized by the DOP, which is calculated as
\begin{equation}
\hbox{DOP}(z)=\frac{\sqrt{\langle S_1^{(s)}(z)\rangle^2+\langle S_2^{(s)}(z)\rangle^2+\langle S_3^{(s)}(z)\rangle^2}}{\langle S_0^{(s)}(z)\rangle}\, .
\label{20}
\end{equation}
As usual, a DOP equal to unity means that light if perfectly polarized, a DOP equal to zero indicates that the light beam is unpolarized, while intermediate
values stand for a partially polarized beam. We say that the Raman polarizer perfoms its function properly when DOP becomes close to unity. Introducing
gain $G$ as $G\equiv \langle S_0^{(s)}(L)\rangle /S_0^{(s)}(0)$ we get $G=\frac{1}{2}\left[ 1+\exp (gPL) \right]$ and for the DOP:
\begin{equation}
\hbox{DOP}=1-G^{-1}\, .
\label{21}
\end{equation}
The higher the gain, the larger the DOP.  Already $20$~dB gain is enough to get a DOP as high as $0.99$. 

A short comment is in order on how one should interpret the averaging procedure, expressed by $\langle\dots\rangle$.  There are two possible 
situations. On the one hand, we can vary the SOP of the signal beam in time, then $\langle\dots\rangle = \langle\dots\rangle_T=T^{-1}\int_0^T \dots \, dt$, 
where $T$ is the period of time, sufficiently long to get correct statistical averaging.  
$\langle S_1^{(s)}\rangle_T = \langle S_2^{(s)}\rangle_T = \langle S_3^{(s)}\rangle_T = 0$ means that we are dealing with unpolarized light. On the other
hand, we can imagine an experiment with an ensemble of beams. Then, $\langle\dots\rangle = \langle\dots\rangle_e$ means ensemble average over all
these beams. If the SOPs of all beams from the ensemble randomly or uniformly cover the Poincar\'{e}  sphere, then, similarly to the time average, we
get $\langle S_1^{(s)}\rangle_e = \langle S_2^{(s)}\rangle_e = \langle S_3^{(s)}\rangle_e = 0$. In this situation we say that we are dealing with an
ensemble of scrambled beams.  In a case where the time average gives the same statistical information as the ensemble average, we refer to such
system as an ergodic one. The Raman polarizers considered here are obviously ergodic systems, simply because time does not enter the equations of
motion explicitely. Therefore, our analysis is valid for the scrambled beams approach as well as for time averaging.

Another important quantity which characterizes a Raman polarizer is the alignment parameter $A_{\uparrow\uparrow}$, defined as the cosine of the
angle between the output signal SOP and the output pump SOP:
\begin{equation}
A_{\uparrow\uparrow}=\frac{\langle S_1^{(s)}S_1^{(p)}+S_2^{(s)}S_2^{(p)}+S_3^{(s)}S_3^{(p)}\rangle}{\langle S_0^{(s)}\rangle S_0^{(p)}}\, .
\label{22}
\end{equation}
The closer the alignment parameter to unity, the better the alignment of the output signal and pump Stokes vectors. Using solutions in 
Eqs.~(\ref{16})-(\ref{19}) we get 
\begin{equation}
A_{\uparrow\uparrow}=\frac{\langle S_1^{(s)}(L)\rangle}{\langle S_0^{(s)}\rangle}=1-G^{-1}
\label{23}
\end{equation}
for the value of the alignment parameter at the fiber output. Although this value coincides with the value of DOP, see Eq.~(\ref{21}), these two quantities
have different physical meanings. For a statistical ensemble of beams, the alignment parameter shows the average direction of the signal Stokes vector
on the Poincar\'{e} sphere, while the DOP measures the breadth of the spot traced by the tips of the signal Stokes vectors around this average direction.

Yet another quantity of interest is the measure of the polarization-dependent gain (PDG). It is exactly the PDG which is at the heart of a Raman polarizer.
Different SOPs of the signal beam experience different amplifications. The signal beam with a SOP parallel  to  the pump Stokes vector is amplified
most efficiently, while the orthogonal polarization experiences no gain. Indeed, as it follows from the solution in Eq.~(\ref{13}), $G_{\max} =\exp (gPL)$
and $G_{\min}=1$. We introduce the PDG parameter $\Delta$ as $\Delta =G_{\max} -G_{\min}$, and get for the ideal  Raman polarizer $\Delta =2(G-1)$.
The PDG parameter aquires high values for a high-gain Raman polarizer. Note that for an ``ideal Raman amplifier" (an amplifier, which is perfectly
described by the  {\it scalar} theory, or in other words, the amplifier, which works deeply in the diffusion limit) $\Delta =0$.

The high value of the PDG parameter points out that along with the desirable property of strong re-polarization of the signal beam, this device is
characterized by a high level of unwanted relative intensity noise (RIN). By varying the signal SOP at the input we get pronounced variations of the
intensity at the output, even if the incoming beam had a steady intensity in time.  In order to characterize the output power fluctuations, let us calculate 
the variance
\begin{equation}
\sigma_s^2 = \frac{ \langle S_0^2(L)\rangle } { \langle S_0(L)\rangle^2 }-1\, .
\label{24}
\end{equation}  
Using solution in Eq.~(\ref{12})  we get
\begin{equation}
\sigma_s^2 = (1-G^{-1})^2/3\, .
\label{25}
\end{equation}  
For large $G$, $\sigma_s\approx 3^{-1/2}\approx 58${\%}.This level of RIN may be detrimental for some optical devices, particularly nonlinear ones. 
Note that an ideal Raman amplifier is characterized by $\sigma_s=0$, i.e., by zero RIN, thanks to the efficient polarization scrambling which is provided 
by PMD. The price to be paid is the totally stochastic signal SOP at the output fiber end. 

A reasonable question to ask is whether it is possible for a Raman polarizer to keep the useful property of re-polarization and at the same time to
suppress RIN down to an acceptable level. The answer is positive. One possible way to combat the RIN and at the same time keep the property of
re-polarization is to use the Raman polarizer in the depleted-pump regime, Ref.~\cite{ourPTL}. In this saturation regime all input SOPs are amplified to
approximately the same level of intensity, actually up to $S_0^{(s)}(L)\approx P$. Strictly speaking, only one signal SOP (the one which is perfectly
orthogonal to the pump SOP) is not amplified at all. However, the numerous imperfections of any practical realization of a Raman polarizer, including
residual PMD, may prevent the observation of such a singular behaviour.  

So far we have analyzed the main statistical properties of an ideal Raman polarizer operating in the undepleted-pump regime. If necessary, any
other statistical quantity of interest can be obtained from the exact analytical solutions given in Eqs.~(\ref{12})-(\ref{15}). In a similar manner, one can
characterize the re-polarization of partially polarized beams. The final quantity which we would like to comment on is the mean gain of an ideal Raman
polarizer. It is well known, that the gain of an ideal Raman amplifier is equal to $g/2$. The reason is that in the course of propagation the signal SOP
rotates quickly around the pump SOP, and on average ``feels" the arithmetic mean of the maximal gain ($g$) when it is parallel to the pump SOP, and
minimal gain ($0$) when it is orthogonal, yielding $g/2$ on average. In terms of available gain, an ideal Raman polarizer performs much better. As can
be seen from Eq.~(\ref{16}), for large values of $G$, $G\approx \exp (gPL-\ln 2)$, so that the gain coefficient is almost twice larger. This property makes 
Raman polarizers very efficient Raman amplifiers as well.

\section{Counter-propagating Raman polarizers}
So far, we have been dealing only with the co-propagating geometry. In this geometry, the pump SOP stochastically changes along the fiber,
and its output SOP depends on the particular realization of the birefringence stochasticity in the chosen fiber span. Moreover, the stochasticity 
changes with time, as a result of variation of the environmental conditions. Therefore the trapping of signal's SOP to pump's SOP does not 
garantee the absence of fluctuations of signal's SOP at the output, even though these fluctuatons closely follow the time-varying pump SOP. 
In other words, the co-propagating Raman polarizer provides the trapping effect in the stochastic frame, but does not garantee the SOP 
stabilization in the laborotary frame.

The desirable stabilization in the laboratory frame can be achieved by implementing a counter-propagating geometry, Refs.~\cite{our2,padova2}.
Since the signal's SOP is
attracted toward the instantaneous position of pump's Stokes vector, this alignment holds also at the output end of the fiber. The output pump SOP is
defined solely by the source, and as such it is supposed to be well defined and deterministic. In this respect the counter-propagating geometry is
preferrable. As regarding the theory, one can repeat derivations with the opposite sign of $z$-derivative in the equation governing evolution of the pump
beam. As shown in Ref.~\cite{our2}, this reversing of the sign brings some changes in the components of the XPolM and Raman tensors. They become
\begin{eqnarray}
&&  J^{counter}_x=-\frac{8}{9}\hbox{diag}(1,\, -1,\, 1)\exp (-z/L_d)\, ,
\label{26}\\
&&  J^{counter}_R=\frac{1}{3}\hbox{diag}(1,\, -1,\, 1)\exp (-z/L_d)\, .
\label{27}
\end{eqnarray}
The presence of the factor $\frac{1}{3}$ in front of the Raman tensor immediately leads us to the conclusion that the counter-propagating Raman 
polarizer is significantly less effective in re-polarization than its co-propagating analog. In order to get similar performances we need either to increase the
pump power or lengthen the fiber, or both. Let us evaluate the performance of this device.

First of all, we start with the solving the equation of motion (\ref{8}) in the undepleted-pump regime. We get
\begin{eqnarray}
&& S_0^{(s)}(z)=\frac{1}{2}\left[ S_{0}^{(s)}(0)-S_{1}^{(s)}(0)\right]\hbox{e}^{\frac{1}{3}gPz}
\nonumber\\
&& +\frac{1}{2}\left[ S_{0}^{(s)}(0)+S_{1}^{(s)}(0)\right]\hbox{e}^{\frac{2}{3}gPz}\, ,
\label{28}\\
&& S_1^{(s)}(z)=-\frac{1}{2}\left[ S_{0}^{(s)}(0)-S_{1}^{(s)}(0)\right]\hbox{e}^{\frac{1}{3}gPz}
\nonumber\\
&& +\frac{1}{2}\left[ S_{0}^{(s)}(0)+S_{1}^{(s)}(0)\right]\hbox{e}^{\frac{2}{3}gPz}\, ,
\label{29}\\
&& S_2^{(s)}(z)=\Big[ S_{2}^{(s)}(0)\cos (\bar\gamma Pz)
\nonumber\\
&& -S_3^{(s)}(0)\sin (\bar\gamma Pz)\Big]\hbox{e}^{\frac{1}{2}gPz}\, ,
\label{30}\\
&& S_3^{(s)}(z)=\Big[ S_{2}^{(s)}(0)\sin (\bar\gamma Pz)
\nonumber\\
&& +S_3^{(s)}(0)\cos (\bar\gamma Pz)\Big]\hbox{e}^{\frac{1}{2}gPz}\, .
\label{31}
\end{eqnarray}
We immediately observe that the difference in amplification coefficients of the first Stokes component and the second (and third) Stokes component
is given by $\frac{2}{3}g-\frac{1}{2}g$. The contrast is much weaker than for the co-propagating case, where we had  $g-\frac{1}{2}g$. The average gain
of the counter-propagating Raman polarizer is 
\begin{equation}
G=\frac{1}{2}\left(\hbox{e}^{\frac{2}{3}gPL}+\hbox{e}^{\frac{1}{3}gPL}\right)\, ,
\label{32}
\end{equation}
which is significantly smaller than for a Raman polarizer operating in the co-propagating configuration, although it is still larger than for an ideal Raman 
amplifier. For the same value of the product $PL$, the DOP for the conter-propagating configuration is also smaller:
\begin{eqnarray}
\hbox{DOP} & = & 1-2\left(\hbox{e}^{\frac{1}{3}gPL}+1\right)^{-1}
\nonumber\\
& \approx & 1-2\hbox{e}^{-\frac{1}{3}gPL}\,\hbox{}\, (\hbox{for}\, gPL\gg 1)
\nonumber\\
& \approx & 1-\sqrt{2}G^{-1/2}\, .
\label{33}
\end{eqnarray} 
For $G=20$~dB in the co-propagating case the DOP was as high as $99${\%}, while in the counter-propagating configuration it is only $86${\%}.

\begin{figure}
\begin{center}
\includegraphics[scale=0.5]{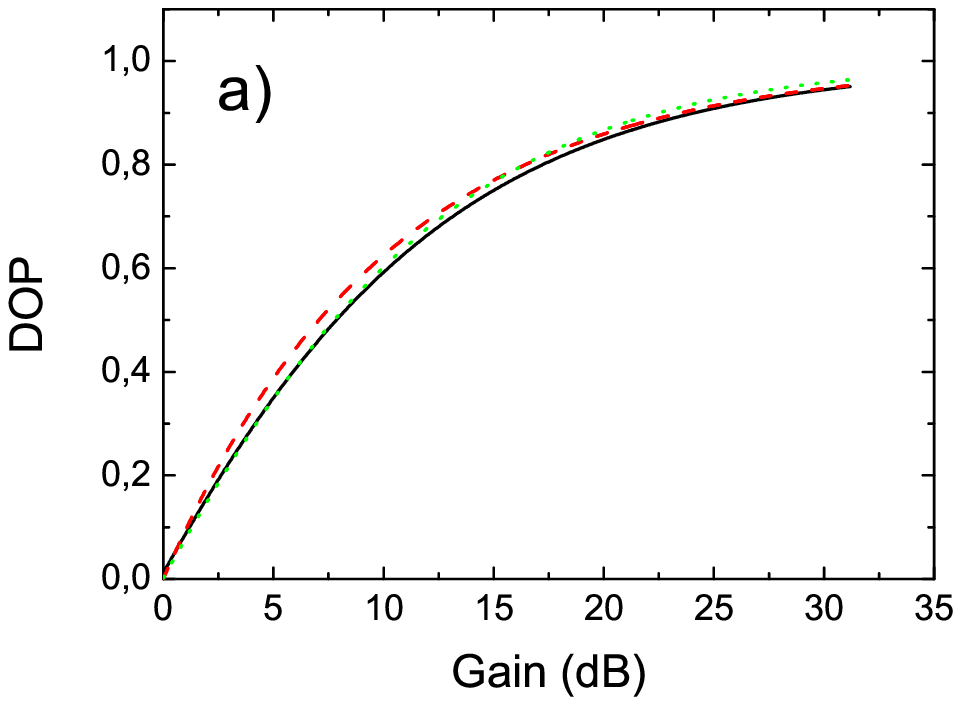}
\includegraphics[scale=0.5]{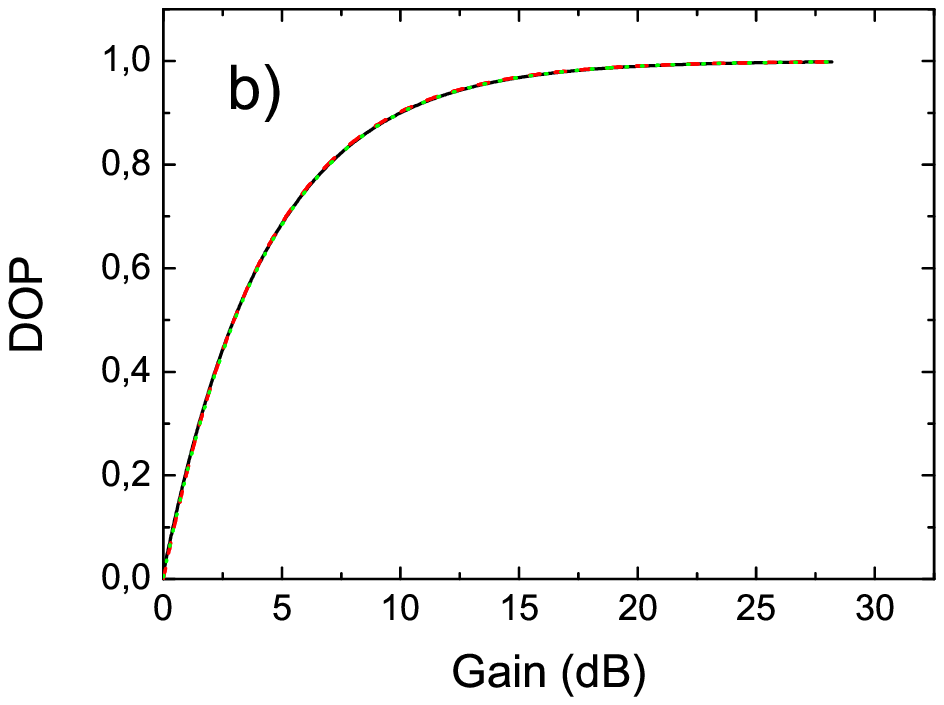}
\end{center}
\caption{DOP versus gain $G$. Graphical comparison of the results obtained with formulae (\ref{33}) (black solid); results obtained with empirical
formulae (\ref{34}) (green dotted); and results based on the direct numerical solution of Eq.~(\ref{3}) with XPolM and Raman tensors in the form of
Eqs.~(\ref{26})-(\ref{27})  (red dashed). Parameters are: a) $L_c=1$~m, $L_B=45$~m, $P=8$~W,  $D_p=0.005$~ps$/\sqrt{\hbox{km}}$, and
$L_d=17.5$~km, $L$ varies from $0$~km to $2.5$~km, and $\Gamma =10.2$; b) $L_c=10$~m, $L_B=3500$~m, $P=8$~W, 
$D_p=0.0002$~ps$/\sqrt{\hbox{km}}$, and $L_d=10914$~km, $L$ varies from $0$~km to $1.5$~km, and $\Gamma =4.3$.
}
\label{ris2}
\end{figure}

It is instructive to compare our model of ideal Raman polarizer with full-scale numerical simulations of the underlying stochastic equations presented
in \cite{padova2}, where the empirical formula:
\begin{equation}
\hbox{DOP}=1-\hbox{e}^{-G_{dB}/\Gamma}\, ,
\label{34}
\end{equation}
connecting the DOP with the gain was suggested and tested numerically. Here $G_{dB}=10\log_{10}G$  and $\Gamma\approx 10.2$ for the considered
range of PMD coefficients. The graphical comparison of the results obtained with formula (\ref{33}) on one hand, and the results plotted according to 
the empirical formula (\ref{34}) on the other hand, is shown in Fig.~\ref{ris2}(a). The fit is good. On the same plot we have also shown the results based
on the direct numerical solution of Eq.~(\ref{3}) with XPolM and Raman tensors in the form of Eqs.~(\ref{26})-(\ref{27}). Note that we did not use any 
fitting parameter in this cross-comparison. 

The alignment parameter for the counter-propagating geometry is different from the co-propagating case. Because of the change of the sign in front of
the second  element of the Raman tensor, see Eq.~(\ref{27}), the alignment parameter is now
\begin{equation}
A_{\uparrow\downarrow}=\frac{\langle S_1^{(s)}S_1^{(p)}-S_2^{(s)}S_2^{(p)}+S_3^{(s)}S_3^{(p)}\rangle}{\langle S_0^{(s)}\rangle S_0^{(p)}}\, .
\label{35}
\end{equation}
For input unpolarized light, the alignment parameter coincides with the DOP, namely,
\begin{eqnarray}
A_{\uparrow\downarrow} & = & 1-2\left(\hbox{e}^{\frac{1}{3}gPL}+1\right)^{-1}
\nonumber\\
& \approx & 1-2\hbox{e}^{-\frac{1}{3}gPL}\,\hbox{}\, (\hbox{for}\, gPL\gg 1)
\nonumber\\
& \approx & 1-\sqrt{2}G^{-1/2}\, .
\label{36}
\end{eqnarray} 

The PDG parameter $\Delta = G_{\max}-G_{\min}$ is easily calculated, resulting in
\begin{equation}
\Delta =\frac{1}{2}\left(\hbox{e}^{\frac{2}{3}gPL} - \hbox{e}^{\frac{1}{3}gPL}\right) =\frac{1}{2}\left( 1+2G-\sqrt{1+8G}\right)\, .
\label{37}
\end{equation}
Its value is considerably less in the co-propagating configuration, particularly for moderate values of gain. This observation again points to the relatively 
poorer performances of the counter-propagating Raman polarizer. At the same time, the RIN is expected to have a lower level. In order to demonstrate
this, let us evaluate the variance of the signal intensity.  Formula (\ref{24}) and solution (\ref{28}) yield
\begin{equation}
\sigma_s^2  = \frac{1}{3} \left[ 1 - 2\left( \hbox{e}^{\frac{1}{3}gPL}+1 \right)^{-1}\right]^2\, .
\label{38}
\end{equation}

Before concluding this section, one remark is in order about the applicability domain of these results. SPolM, XPolM and Raman tensors given by
Eqs.~(\ref{4}), (\ref{5}), (\ref{6}), (\ref{26}), and (\ref{27}) were calculated in the limit
\begin{equation}
L\gg L_{bire}\equiv\frac{L_B^2(\omega_p)}{8\pi^2 L_c}\, .
\label{39}
\end{equation}
This inequality holds for all practical situations. Thus, for $L_c$ as small as $1$~m and $L_{B}$ as large as $100$~m we get $L_{bire}$ as short as 
$127$~m. Fiber-optic Raman amplifiers are always longer than $1$~km, and therefore inequality (\ref{39}) is not violated.
However, if for some reason inequality (\ref{39}) is violated, for instance for extremely low birefringent 
fibers, the analysis given above must be corrected. Thus, in the limit $L_B(\omega_p)\to\infty$, the tensors of interest take the following form:
\begin{eqnarray}
J_s^{counter}=\hbox{diag}(-1,\, 1,\, -1)\, ,
\label{40}\\
J_x^{counter}=\frac{4}{3}\hbox{diag}(-1,\, 0,\, 1)\, ,
\label{41}\\
J_R^{counter}=\hbox{diag}(1,\, 1,\, 1)\, .
\label{42}
\end{eqnarray}
Fig.~\ref{ris2}(b) shows the dependence of DOP on Raman polarizer gain for this case. Although the performance of the Raman polarizer in this limit is 
very good, we will not evaluate it here explicitely because of its little practical interest.

\section{Conclusion}
We have studied the effect of trapping of the state of polarization of a signal beam by a pump beam in the model of a Raman polarizer. We have 
introduced the notion of the ideal Raman polarizer and quantified its performance it terms of gain, degree of polarization, polarization-dependent gain 
parameter, alignment parameter, and RIN characteristics. We have studied two different geometries: co - and counter-propagating configurations, and 
identified their pros and contras. Possible applications of Raman  polarizers include their potential use in telecom-related signal processing, where the 
need of transforming an unpolarized light to a polarized one is necessary in order to provide an interface between the telecom link and post-processing
polarization-sensitive devices (based, for instance, on nonlinear crystals). 


\end{document}